\documentclass[aip,jap,reprint,amsmath,amssymb]{revtex4-1}
\usepackage{graphicx}% Include figure files
\usepackage{dcolumn}% Align table columns on decimal point
\usepackage{bm}% bold math
\usepackage{verbatim}

\begin{document}

%\begin{comment}
\title{Spin Seebeck devices using local on-chip heating}

\author{Stephen M. Wu}
	\email{swu@anl.gov}
\affiliation{%
Materials Science Division, Argonne National Laboratory, Argonne, Illinois 60439, USA
}%

\author{Frank Y. Fradin}
\affiliation{%
Materials Science Division, Argonne National Laboratory, Argonne, Illinois 60439, USA
}%

\author{Jason Hoffman}
\affiliation{%
Materials Science Division, Argonne National Laboratory, Argonne, Illinois 60439, USA
}%

\author{Axel Hoffmann}
\affiliation{%
Materials Science Division, Argonne National Laboratory, Argonne, Illinois 60439, USA
}%

\author{Anand Bhattacharya}%

\affiliation{%
Materials Science Division, Argonne National Laboratory, Argonne, Illinois 60439, USA
}%

\date{\today}% It is always \today, today,

\begin{abstract}
A micro-patterned spin Seebeck device is fabricated using an on-chip heater. Current is driven through a Au heater layer electrically isolated from a bilayer consisting of Fe$_3$O$_4$ (insulating ferrimagnet) and a spin detector layer. It is shown that through this method it is possible to measure the longitudinal spin Seebeck effect (SSE) for small area magnetic devices, equivalent to traditional macroscopic SSE experiments. Using a lock-in detection technique it is possible to more sensitively characterize both the SSE and the anomalous Nernst effect (ANE), as well as the inverse spin Hall effect in various spin detector materials. By using the spin detector layer as a thermometer, we can obtain a value for the temperature gradient across the device. These results are well matched to values obtained through electromagnetic/thermal modeling of the device structure and with large area spin Seebeck measurements.
\end{abstract}

\maketitle

The spin Seebeck effect (SSE) has been widely studied due to the implications it has on the generation of pure spin currents\cite{uchida2008observation,uchida2010observation}. In the SSE, applying a thermal gradient across a magnetic insulator generates a pure spin current that flows into an adjacent material without any charge current\cite{uchida2010spin}. Experiments involving the SSE have taken many forms, but the canonical method has been to use large area ferromagnetic insulators on the order of several millimeters combined with Peltier elements to generate a thermal gradient. There are several disadvantages to this technique including being sensitive to material non-uniformity, being confined to using large samples, and being limited by complicated experimental setups.

Here we introduce a micro-patterned spin Seebeck device to solve many of these issues. We measure a microscale SSE using confined local on-chip heating, which is electrically separated from both the spin current source and spin current detector. Using this method it is possible to sensitively examine local magnetization, local spin current, and local spin to charge conversion in a simple and scalable device structure. Since the devices are small, it is possible to easily integrate them into conventional cryostat systems. While there are alternative small-area spin Seebeck techniques such as laser heating \cite{laserheating} or current induced heating using the spin detector layer\cite{schreier2013current}, each has its disadvantages. Laser heating involves additional experimental setup, making it harder to integrate into existing device measurement setups, and current induced heating using the spin detector layer generates several undesired conventional charge transport effects that must be separated to extract the SSE. 

\begin{figure}[b]
\includegraphics[width=3.4in,trim =2.5in 1in 2.25in 1in,clip=true]{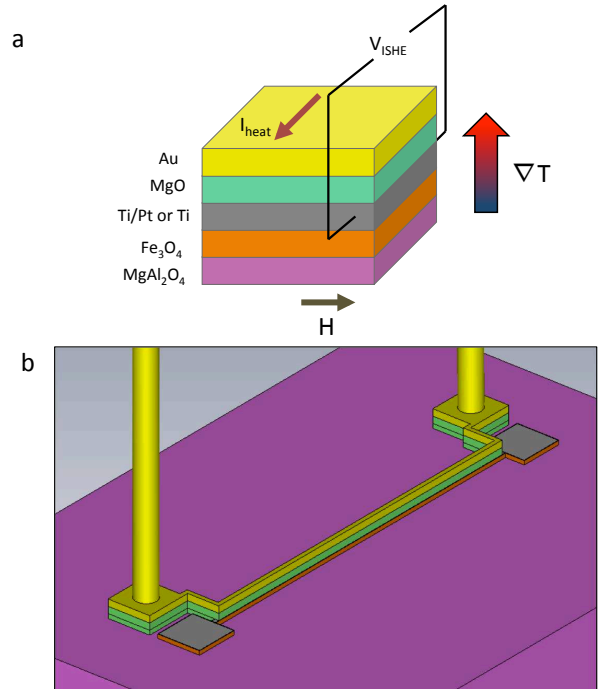}
\caption{\label{fig1} A schematic of a typical spin Seebeck device is presented in (a).  A model of the actual device structure used in experiments and simulations is presented in (b). The pillars represent wirebonds made to the device, the wirebonds on the second set of pads are hidden for viewing clarity.  }
\end{figure}	 

In this work, our devices consist of Fe$_3$O$_4$ (60 nm)/X/MgO (100 nm)/Au (100 nm) on MgAl$_2$O$_4$ substrates, where X is either Ti (15 nm), Ti (1.5 nm)/Pt (5 nm) or absent (Fig. 1a). On-chip heating is provided by the Au layer, which is electrically isolated from the rest of the device by a layer of MgO. Fe$_3$O$_4$ serves as the spin current source, while X serves as a spin detector layer that responds to spin current through the inverse spin Hall effect (ISHE). 

Fe$_3$O$_4$ is grown on MgAl$_2$O$_4$ (100) substrates using ozone assisted oxide molecular beam epitaxy (MBE), described elsewhere \cite{liu2013non}. The film is then patterned into 10 $\mu$m x 800 $\mu$m strips through standard photolithography and liquid nitrogen cooled Ar$^+$ ion milling. Liquid nitrogen cooling is used to limit defects introduced by the ion milling process. Using a lift-off process with standard photolithography and electron beam evaporation, a layer of Ti/Pt or Pt is then deposited onto the Fe$_3$O$_4$ strip. Finally, using the same lift-off process the MgO/Au heater layer is patterned onto the device, with separate contacts to the side. The finished device is shown in a 3D model in Fig. 1b. 

The response in the spin current detector layer is  $\vec{E}_{ISHE}\propto \vec{J_S} \times \hat{\sigma}$, where $\vec{J_S}$ is the injected spin current, and $\hat{\sigma}$ is the unit vector of the spin. Because the spin current injected into the adjacent spin detector layer is directly proportional to the thermal gradient and the magnetization, the total response is $\vec{E}_{SSE}\propto  \nabla T \times \vec{M}$. We are not limited to using DC techniques to detect the ISHE signal from the spin detector layer. By using AC lock-in detection, the sensitivity of the measurement can be greatly increased. Since the power generated through Joule heating $P=I_{heater}^2R \propto \nabla T$, the voltage measured will also be $\propto I_{heater}^2$. If an AC signal is used to excite the heater $I_{heater}\propto sin(\omega t)$, then the measured voltage $V\propto \frac{1}{2}(1 -cos(2\omega t))$. By lock-in detecting the 90$^{\circ}$ out-of-phase component at the $2\omega$ frequency it is possible to detect the spin Seebeck effect with much higher sensitivity because parasitic effects occur at the $\omega$ frequency. By using this technique and ignoring the constant term, it is possible to detect signals as small as 2 nV depending on the integration period of the lock-in amplifier. 

The sample is mounted onto a standard circuit board using silver paint, and contact to the device is made using wirebonds before it is loaded into a  Quantum Design Physical Properties Measurement System (PPMS) cryostat for temperature and magnetic field control. Experiments performed on three different device stacks are presented in Fig. 2, each with a different spin current detector layer (Ti, Ti/Pt, or Fe$_3$O$_4$ alone). In each experiment a constant 5 V$_{pp}$ signal at 99 Hz is applied across the Au heater and a 60 ohm load resistor. Since the resistance of the heater layer changes with temperature, the power applied using this measurement also changes from 17.2 mW$_{rms}$ at 300 K  to 21.9 mW$_{rms}$ at 15 K. At each temperature the voltage across the device is measured with respect to an in-plane magnetic field applied transverse to $\vec{E}_{SSE}$.  A typical curve is presented in the inset of Fig. 2. The magnitude of the voltage response, as defined as the voltage difference at zero applied magnetic field for the two different magnetization states of Fe$_3$O$_4$,is shown against temperature in Fig. 2.

\begin{figure}[ht]
\includegraphics[width=3.4in,trim =0in 0.25in 0in 0.75in,clip=true]{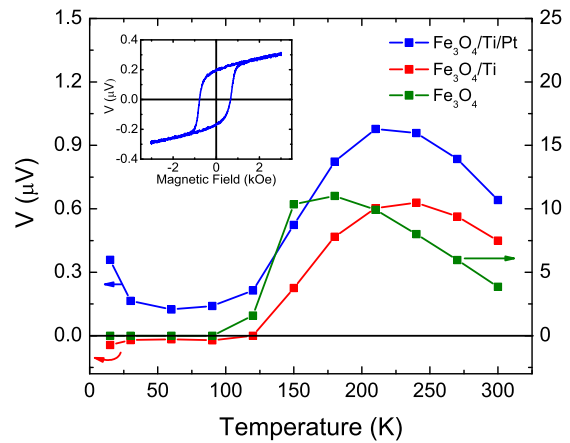}
\caption{\label{fig2} The magnitude of the spin Seebeck/anomalous Nernst response with respect to temperature. Devices with spin detector layers of Ti (15 nm), Ti(1.5 nm)/Pt (5 nm), and a control device with Fe$_3$O$_4$ are presented. The inset curve is an example of a spin Seebeck signal with respect to magnetic field for a Ti/Pt device at 15 K. The linear contribution to this voltage response is likely due to the ordinary Nernst effect from the spin detector materials. }
\end{figure}	 

Fe$_3$O$_4$ is only insulating below a well known metal-insulator transition at 120 K known as the Verwey transition. At temperatures above the Verwey transition there is a contribution from both the anomalous Nernst effect from conducting Fe$_3$O$_4$, and the ISHE due to the SSE in Ti or Ti/Pt. At these temperatures it is not possible to separate the two effects \cite{wu2014unambiguous}. However, below the Verwey transition ,there is a recovery in both the Ti/Pt and the Ti devices due to the SSE, while the signal in the Fe$_3$O$_4$ control device goes to zero. The 1.5 nm Ti spacer layer in between the Fe$_3$O$_4$ and the Pt in the Ti/Pt device serves to remove any effects from proximity magnetism \cite{huang2012transport,kikkawa2013longitudinal,geprags2012investigation}. The large difference in magnitudes between the Fe$_3$O$_4$ control device and devices with spin detector layers are due to the large difference in resistivities between the different films. The low resistance spin detector layers essentially act as a shunt that reduces the total voltage response \cite{wu2014unambiguous,ramos2013observation}. At the same time, the difference in the magnitudes between the Ti and the Ti/Pt devices is explained by the individual spin Hall angles of the two materials. The spin Hall angle in Ti is small and negative, as predicted theoretically through tight binding calculations \cite{tanakaintrinsicspinhall} and experimentally observed \cite{uchida2014longitudinaltitanium}, while Pt has been the long standing canonical standard for ISHE experiments due to its large ISHE response \cite{hoffmanspinhallmetals}. Our measurements reproduce these results well, showing that our method is equivalent to large area macroscopic measurements. 

To look further in detail into the heat flow in our devices, thermal simulations were performed on the device model presented in Fig. 1b, using the Computer Simulation Technology (CST) Studio Suite software modeling package. Values for material resistivity were taken from measured values, while values for thermal conductivity and heat capacity were taken from standard sources in literature \cite{slack1962thermal,heatcapcityshepherd}. Since the device sits on a large substrate that is mounted using silver paint onto a standard PPMS rotator circuit board  solidly heat sinked to the PPMS, the back side of the substrate is assumed to be held at the bath temperature. The same condition is assumed of the four wirebond contacts to the device, since each wirebond is also connected directly to the circuit board. The DC heater current chosen for this simulation was equivalent to the maximum peak current in our AC measurements, 24.5 mA at 300 K and 30.1 mA at 15 K. The resulting heater power loss density is presented in Fig. 3a for a Ti/Pt device at 300 K. This shows that all the heater power is constrained to the narrowest and most resistive part of the device where the Fe$_3$O$_4$/Ti/Pt device stack is. The resulting temperature within the device is presented in Fig 3b. The $\Delta$T from the back of the substrate to the top of the heater is 2.226 K at 300 K, and 0.565 K at 15 K. Given this $\Delta$T, the voltage response measured in our devices match well with large area SSE experiments on both thin film and bulk ferromagnets \cite{ramos2013observation,uchida2012thermal}. Since the large area experiments show that there is a linear voltage response to thermal gradient up to $\Delta$T= 20 K, by optimizing the power transferred to our heater our signal to noise ratio could be increased even further by increasing $\Delta$T in our system. The averaged $\Delta$T along the device length, shows the uniformity of the thermal gradient is within 3\% according to simulations. 

Although there is a large $\Delta$T across the entire substrate and device, the temperature drop across just the Fe$_3$O$_4$ layer is not as large and depends its  individual thermal conductivity. Characterizing $\Delta$T across thin film ferromagnets has been an ongoing challenge in SSE experiments both on the macroscale and the microscale since there is no standard method to probe the temperature at both sides of the thin film \cite{wu2014unambiguous}. By applying a constant $\Delta$T across the substrate and the thin film ferromagnet there is no guarantee that the thermal gradient across the thin film will be constant with respect to temperature since the thermal conductivities of the substrate and the thin film will change relative to each other. By using an on-chip heater, we effectively send a constant heat current through the device instead of setting up a constant $\Delta$T. This is analogous to performing a current biased vs. voltage biased electrical measurement. Using this method it may be easier to eliminate substrate based effects since the thermal gradient across the thin film ferromagnet will only depend on the properties of the thin film and not its relative value compared to the substrate. 

\begin{figure}[t]
\includegraphics[width=3.4in,trim =1in 1in 0.75in 0.5in,clip=true]{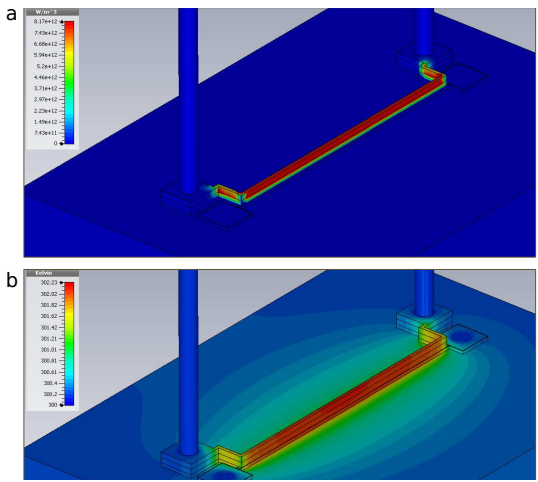}
\caption{\label{fig3} Electromagnetic and thermal simulation results for a Ti/Pt spin Seebeck device. (a) shows the calculated power loss density due to Joule heating in the heater layer, while (b) shows the calculated local temperature within the device that results from heating.  }
\end{figure}	 

Using thermal simulation it is possible to model the out-of-plane thermal gradient across each thin film in the center of our device (Fig. 4). The thermal gradient across just the Fe$_3$O$_4$ layer is 0.298 K/$\mu$m at 300 K and 0.014 K/$\mu$m at 15 K. This difference is consistent with the relative magnitudes of the thermal conductivity of Fe$_3$O$_4$ at 300 and 15 K. The inset of Fig. 4 shows that the much larger temperature drop is across the substrate due to its size relative to the thin film  (550 $\mu$m vs. 0.06 $\mu$m). It also shows that the temperature drop across the substrate is non-uniform due to the heating being localized to the device and not the entire top surface of the chip. Locally at the surface, on the scale of our thin films, the temperature distribution is highly linear. 

\begin{figure}[ht]
\includegraphics[width=3.4in,trim =0in 0.25in 0in 0.75in,clip=true]{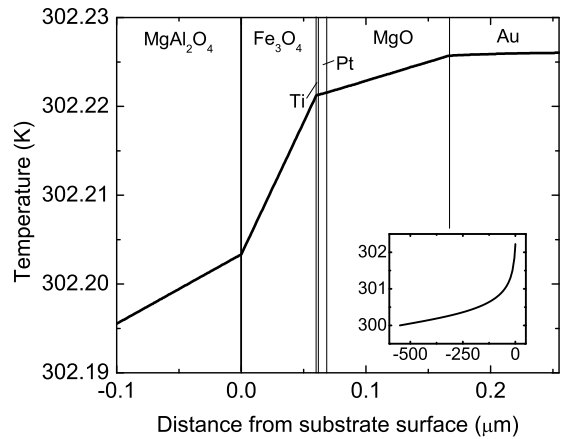}
\caption{\label{fig4} Temperature in the out-of-plane direction at the center of a Ti/Pt spin Seebeck device as calculated from thermal modeling. The inset shows the same temperature distribution at an expanded range through the entire substrate.  }
\end{figure}	 

Using the metallic spin detector layer in our device as a thermometer, it is possible to compare the results of thermal modeling with measured values for $\Delta$T. By characterizing the resistance of the Ti (15 nm) strip in our Fe$_3$O$_4$/Ti spin Seebeck device we have a one-to-one relation between temperature and resistance. By running an AC signal through our on-chip heater and measuring the change in resistance in the Ti strip it is possible to directly measure the change in temperature due to the heater. The resistance of the Ti strip and its derivative with respect to temperature is measured and shown in Fig. 5a-b. Next, the same 99 Hz 5 V$_{pp}$ signal is sent to the heater through a 60 ohm load like in the experiment presented in Fig. 2. The peak to peak amplitude of the change in resistance of the Ti strip is presented in Fig 5c. Finally, the change in temperature is derived from $\Delta T= \frac{\Delta R}{dR/dT}$ (Fig. 5d). Measurements of the thermal time constant of our system revealed two time scales for changes in the resistance of Ti strip at all temperatures. One of which occurs faster than our data acquisition system can resolve (5000 samples/s), and one of which occurs on the scale of seconds. The longer time constant is likely due to heating the entire experimental probe within the cryostat along with the sample. At 99 Hz the only contribution we measure is the short time constant $\Delta$T across just the device. The values at both low and high temperatures show values close to the predicted results of thermal modeling. The remaining differences are likely due to the differences between the values of thermal conductivity taken from literature and our samples. Since our measurements are heat current biased, the results in Fig. 5d scale inversely to measurements of thermal conductivity on MgAl$_2$O$_4$, which seems to be the largest determining factor of $\Delta$T in the device. There is a local increase in $\Delta$T near the Verwey transition in Fig. 5d that cannot be directly explained by thermal conductivity changes in MgAl$_2$O$_4$ or Fe$_3$O$_4$ single crystals \cite{slack1962thermal} alone, but does resemble the discontinuity in Fe$_3$O$_4$ heat capacity at this temperature \cite{heatcapcityshepherd}.

\begin{figure}[t]
\includegraphics[width=3.4in,trim =0in 0in 0in 1.25in,clip=true]{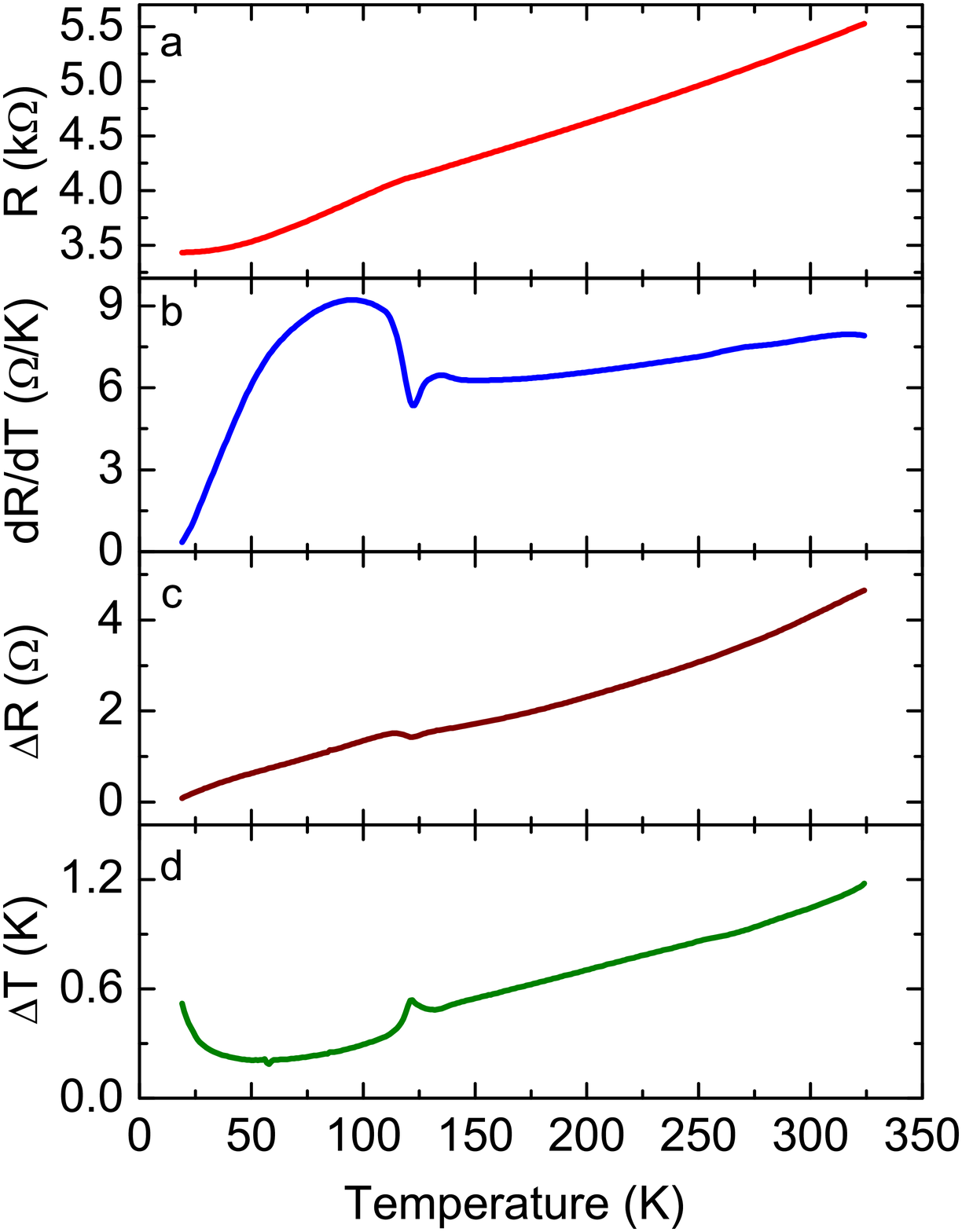}
\caption{\label{fig5} The measured resistance vs. temperature of Fe$_3$O$_4$/Ti strip is presented in (a), along with its derivative with respect to temperature in (b). (c) shows the change in resistance of the Fe$_3$O$_4$/Ti strip upon heating with a constant 5 V$_{pp}$ applied across a 60 ohm load resistor and the heater layer. Using the results from (b) and (c), the temperature difference across the device is calculated with respect to temperature and presented in (d).  }
\end{figure}	 
 
We have introduced a method to measure the spin Seebeck effect using a micropatterned device with an on-chip heater. By using a small scale device it is possible to sensitively measure local magnetization, local spin current, and local spin to charge conversion on the microscale and potentially the nanoscale. By using the spin current detection layer as a thermometer it is possible to extract a equivalent thermal gradient to the constant $\Delta$T measurements in conventional spin Seebeck experiments. These measurements match well with thermal simulations of our device structure. This type of device structure allows for easier access to lower temperature, higher magnetic field, and smaller magnetic materials all of which have been challenging to explore using other methods, which require more custom experimental setup. Through the exploration in these regimes it may be possible to provide further insight into the basic mechanisms behind the SSE, as well as potentially finding other interesting thermal spin transport phenomenon.

\begin{acknowledgments}
All authors acknowledges support of the U.S. Department of Energy (DOE), Office of Science, Basic Energy Sciences (BES), Materials Sciences and Engineering Division. The use of facilities at the Center for Nanoscale Materials, was supported by the U.S. DOE, BES under contract No. DE-AC02-06CH11357.
\end{acknowledgments}

\end{document}